\documentclass[conference]{IEEEtran}
\IEEEoverridecommandlockouts
\usepackage{url}
\usepackage{comment}
\usepackage{graphicx}
\usepackage{float,fancyvrb}
\usepackage{subcaption}
\usepackage{algorithm,algorithmicx,algpseudocode}
\usepackage{amsthm}
\usepackage{amsmath}
\usepackage{mathtools}
\usepackage{multirow}
\usepackage{booktabs}
\usepackage{tikz}
\usepackage{pgfkeys}
\usepackage{pgfplots}
\pgfplotsset{compat=1.12}

\def\BibTeX{{\rm B\kern-.05em{\sc i\kern-.025em b}\kern-.08em
    T\kern-.1667em\lower.7ex\hbox{E}\kern-.125emX}}
\begin{document}

\title{Generating Maximal Configurations and Their Variants Using Code Metrics}

\author{
        \IEEEauthorblockN{Tuba Yavuz}
        \IEEEauthorblockA{\textit{University of Florida}\\
        tuba@ece.ufl.edu}\\
\and
        \IEEEauthorblockN{Chin Khor}
        \IEEEauthorblockA{\textit{Iowa State University}\\
        chinkhor@iastate.edu}\\
\and
        \IEEEauthorblockN{Ken (Yihang) Bai}
        \IEEEauthorblockA{\textit{University of Florida}\\
        baiyihang@ufl.edu}\\
\and
        \IEEEauthorblockN{Robyn Lutz}
        \IEEEauthorblockA{\textit{Iowa State University}\\
        rlutz@iastate.edu}\\        
}

\maketitle

\begin{abstract}

Testing configurable systems continues to be challenging and costly.  Generation of configurations for testing tends to use either techniques based on semantic sampling (e.g., logical formulas over configuration variables, often called {\em presence conditions}) or structural code metrics (e.g., code coverage).   In this paper we describe our hybrid approaches that combine these two kinds of techniques to good effect.   We present new configuration-generation algorithms that leverage constraint solving (SAT and MaxSAT) and configuration fuzzing, and implement our approach in a configuration-generation framework, CONFIZZ.  CONFIZZ both enables the generation of maximal configurations (maximal sets of presence conditions that can be satisfied together) and performs code-metric guided configuration fuzzing.  Results from evaluation on BusyBox, a highly configurable benchmark, show that our MaxSAT-based configuration generation achieves better coverage for several code metrics.  Results also show that, when high coverage of multiple configurations is needed, CONFIZZ’s presence-condition fuzzing outperforms alternatives.    
\end{abstract}


\section{Introduction}
\label{sec:intro}

Configurable systems enable code reuse while supporting variability. 
A configuration space is defined by a set of features and the possible 
values they can take.
The configuration space can be divided into two subspaces:  
the compile-time configuration space and the run-time configuration space.
These configuration spaces require more sophisticated testing techniques to 
deal with the combinatorial explosion of the combined configuration and input space.

This has motivated researchers to investigate the problem from various dimensions, 
including using sampling methods \cite{TLD11,Yilmaz14,SGS15,MKR16,AGZ19,KGS19}; 
extracting configuration constraints \cite{NBK14,LKB18} and 
feature interactions using dynamic analysis \cite{SPF14,NKC16,SMN18}, 
static analysis \cite{KSK19,NNT19}, or both \cite{VJS20,Gentree21};
and exploring variability-aware transformation and analyses \cite{RLJ18,PZP22,SGP22}.
To our knowledge sampling and feature-aware code analysis have been treated 
mostly as complementary with few approaches combining them through 
code coverage \cite{TLD11,AGZ19} or in a loosely coupled way \cite{NNT19}. 
However, for effective bug finding and functional validation, 
these two approaches should be combined using semantic code metrics in addition to structural 
code metrics such as code coverage.

Fuzzing has become a popular testing approach in the software domain. Applications of fuzzing to configurable systems include 
configuration error testing \cite{KUC08,XZH13,ZE15,LLL18} and 
runtime option configuration testing \cite{LAK22,ZKW23,LLL24}. 
There is a need to explore the compile-time configuration space 
using fuzzing and without restricting it to the erroneous 
configuration space.

In this paper we focus on the compile-time configuration space and present a 
hybrid approach to configuration generation that combines 
sampling methods and code metrics.
Our approach leverages existing work on extracting presence conditions, 
which are propositional logic formulas over the configuration variables and 
represent the conditions under which pieces of code get compiled 
into the software \cite{NBKC15}. Presence conditions are used in preprocessor statements such as {\tt \#if} and {\tt \#ifdef}.  
We present several configuration generation algorithms that use constraint solving 
to generate maximal constraints with respect to the code metric being used. 
Our approach has been implemented in a tool called CONFIZZ, which 
can be configured to use any custom code metric. In this paper, we demonstrate 
the use of {\em logical code unit coverage} and 
two additional metrics that measure direct and indirect calls to memory deallocation functions such as the {\tt free} function in glibc.

We explore two types of configuration-generation algorithms here: 1) MaxSAT-based approaches and 2) Fuzzing approaches. MaxSAT-based approaches use a MaxSAT solver to 
find maximal configurations, i.e., the maximal sets of presence conditions that can be satisfied together. 
Fuzzing-based approaches use some representation of the search space (based on configuration variables or presence conditions) to generate variations using 
mutations. The generated candidate sets of presence conditions are validated using a SAT solver 
to filter out the invalid ones. The valid ones are used for further mutations by prioritizing 
those that have higher scores for a given code metric.

To evaluate the effectiveness of our approach, we have applied it to the components of BusyBox, a popular highly configurable system that comes with a default configuration,  
and 
focused on the following research questions: 
{\bf RQ1: What percentage of the presence conditions in each component of the BusyBox benchmarks are covered by the default configuration?}
Our results show that the default configurations are fairly good (averaging around 60\% in our experiments on the editors and coreutil components); however, our maximal configurations can achieve better coverage in individual components. 
{\bf RQ2: How many configurations are needed for each component in the BusyBox benchmark set to cover all the presence conditions in the component?}
Our results show that for most components 100\% presence coverage can be achieved with three configurations, while four configurations sufficed for {\tt ls} and five for {\tt cp}.  
{\bf RQ3: How do the sampling, MaxSAT-based, and fuzzing approaches compare in terms of their maximum coverage scores and  running times?}
CONFIZZ consists of three MaxSAT-based maximal configuration generation approaches and 
three SAT-based configuration fuzzing approaches.
Our results show that Maximal, a MaxSAT-based approach, achieves the best performance both in terms of 
code metric coverage score and runtime overhead, while Maximal Iterative, another MaxSAT-based approach, 
guarantees 100\% coverage of a given code metric with low overhead. However, when 
multiple configurations with high scores are needed, Presence Condition Fuzzing, a SAT-based configuration fuzzing approach
is the method of choice as it provides the highest median scores and outperformed black-box sampling using 2-way or 3-way combinatorial interaction testing.

{\bf RQ4: Do bug-relevant code metrics help generate configurations with relevant bugs?}
Our results show that {\tt free} (the deallocation function) relevant code metrics help CONFIZZ generate configurations 
that are related to double-free/use-after-free bugs. In fact, CONFIZZ found several crashes 
in some of the BusyBox components that have been deemed as exploitable using the maximal configuration it generated. 

Our contributions can be summarized as follows:
\begin{enumerate}
    \item We present a MaxSAT-based maximal configuration generation approach, which we use to implement an approximate algorithm for generating minimal  
    number of configurations. We further partition the configuration space to search for additional maximal configurations. The notion of maximal can be customized using a code metric.
    \item We present 
    configuration fuzzing approaches that use constraint solving and code metrics  to search for maximal configuration sets through mutation. 
    \item We developed a tool, CONFIZZ\footnote{We will release our tool and the benchmarks on github.}
    that 
    implements the MaxSAT-based and fuzzing-based configuration-generation algorithms  
    to support testing of configurable systems.
    \item We 
    applied our approach to BusyBox, a real-world configurable system, and 
    reported new findings that can guide testers.
\end{enumerate}

This paper is organized as follows. In Section \ref{sec:relwork}, we discuss related work.
We present our approach in Section \ref{sec:approach}.
We present the details of evaluating our research questions in Section \ref{sec:eval}.
Finally, we conclude in Section \ref{sec:conclusions}.

\section{Related Work}
\label{sec:relwork}

The goal of existing configuration fuzzing work includes 
configuration error testing \cite{KUC08,XZH13,ZE15,LLL18} and 
runtime option configuration testing \cite{LAK22,ZKW23,LLL24}.
CONFIZZ is complementary to these works as it targets 
compile-time configuration fuzzing. 
It generates maximal variants of compile-time configurations by providing a variety of 
configuration generation techniques using constraint solving and presence condition fuzzing.

ConfErr \cite{KUC08} leverages domain information for misspelling, structural, and semantic errors.
SPEX \cite{XZH13} uses LLVM IR-level analysis to infer constraints on the configuration 
variables and generates deviations from the inferred ones to test for configuration errors. It performs limited mutation (negation) of relational operators that appear in the constraints. However, the IR-level analysis may miss the constraints that are available at the Abstract-Syntax Tree (AST) level, which is leveraged by CONFIZZ.
ConfigDiagDetector \cite{ZE15} mutates configuration settings using random mutations 
and a dictionary. 
ConfVD \cite{LLL18} performs grammar-based mutations.
An empirical study in \cite{LJL21} compares these configuration error injection testing techniques on six popular projects, reporting that specification-based approaches are the most effective with moderate efficiency and great human effort while mutation-based has a trade-off between effectiveness and human effort and limited 
efficiency. 

POWER \cite{LAK22} is a program option-aware fuzzer that explores the space of 
runtime options and augments program fuzzing with a preliminary option fuzzing state that 
mutates program options and uses an option configuration relevance metric to choose the 
seed option configurations for actual software fuzzing. 
The configuration relevance metric for two option configurations are measured using the relevance of functions that get executed for each of the option configuration. 
POWER \cite{LAK22} and the distance metric used are complementary to our 
compile-time configuration space exploration and static code metrics.
ConfigFuzz \cite{ZKW23} fuzzes runtime option configurations along with the inputs, which allows utilization of the mutations and the edge coverage metric 
that are built into the underlying software fuzzer for the runtime configuration space. 
ECFuzz \cite{LLL24} optimizes configuration testing of cloud systems by using  configuration dependency specific mutations and leveraging unit tests to eliminate invalid runtime option configurations.

%Toward detecting inconsistencies between a feature model and the software that implements it, such as  dead features or no valid configuration possible, Le et al. generate test cases to surface defects in the software's feature model \cite{Le21, Directdebug21}. 
%Robyn: they're looking for problems in the FM, not the code

Combinatorial Interaction Testing (CIT) techniques \cite{Yilmaz14} have been applied to the testing of both software product lines and highly configurable systems. CIT identifies a subset of features to be considered in combination to achieve, most commonly, pairwise coverage of potential interactions \cite{Lopez15}. It is a form of sampling, informed by constraints on allowable feature combinations, that reduces the number of tests needed.  The Advanced Combinatorial Testing System (ACTS) \cite{Kuhn2010}, \cite{acts32}, which we use in our experiments, is a widely used Combinatorial Interaction Testing (CIT) tool for generating test sets to cover all {\em t}-way combinations of features, where {\em t} specifies the degree of interactions. We use ACTS to present a variation of CIT that labels the generated 
configurations with code metric scores.

 \cite{TLD11} presents an approximation algorithm by mapping the problem of generating the minimal set of configurations to a graph coloring problem. Our encoding of the problem as a minimum set covering 
 problem allows us to use MaxSAT solvers and leverage the recent advances in SAT solving.

\section{Approach}
\label{sec:approach}

\begin{figure*}[th!]
    \centering
    \includegraphics[width=18cm,trim={0cm 3cm 0cm 2cm},clip]{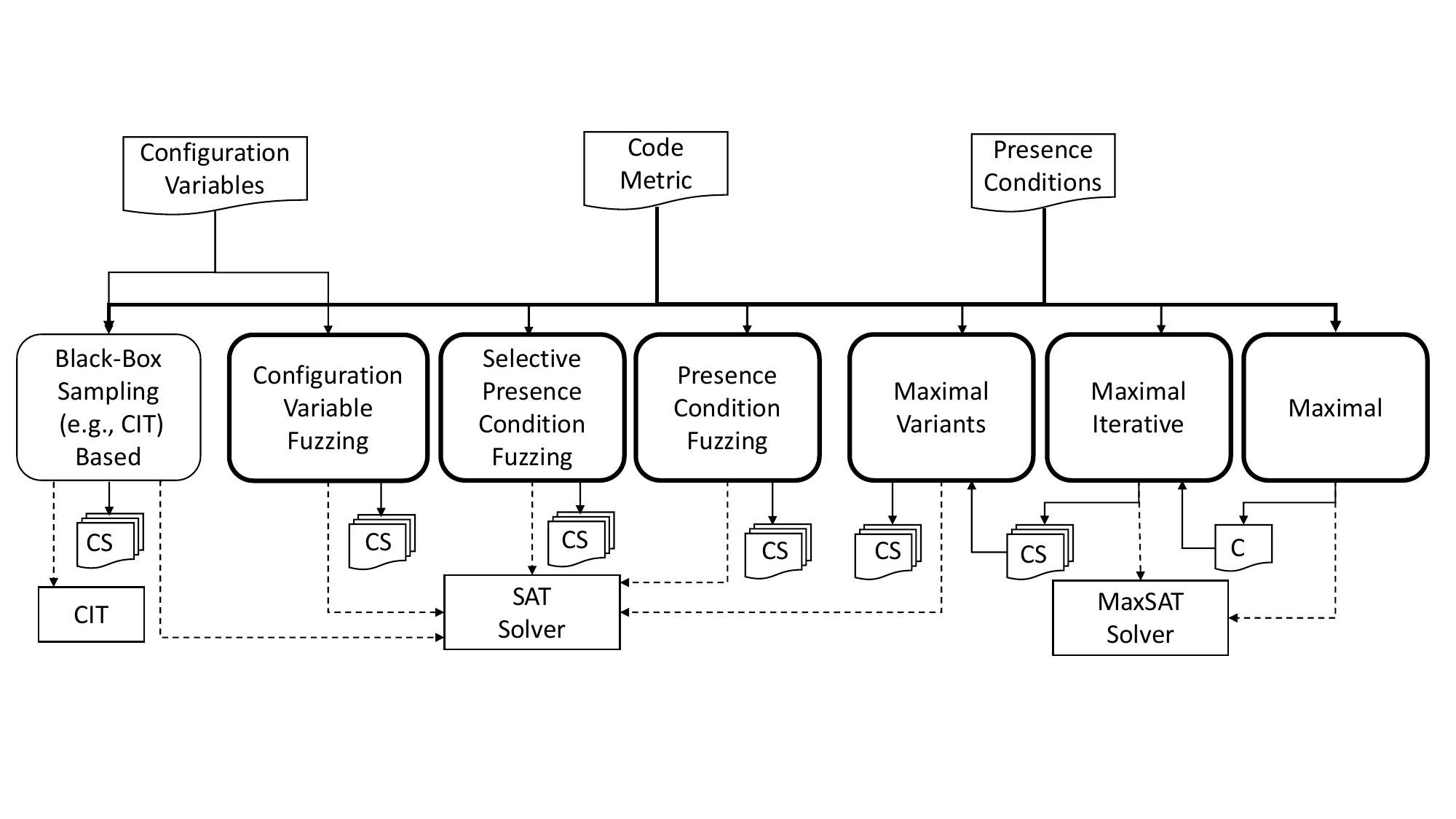}
    \caption{Various configuration generation approaches implemented in {\bf CONFIZZ} are shown with thick rounded rectangles. Solid arrows denote data flow and dashed arrows denote control flow. {\bf C} and {\bf CS} denote a single configuration and a set of configurations, respectively. }
    \label{fig:overview}
\end{figure*}

In this section, we present the details of our approach.
Figure \ref{fig:overview} illustrates the architecture of {\bf CONFIZZ}, which accepts three types of inputs: the configuration variables, the presence conditions in the form 
of propositional logic formula, and a code metric, which maps each presence condition 
to a non-negative score. 
The configuration generation approaches implemented in CONFIZZ are shown using the thick 
rounded rectangles.
The code metric can be statement coverage or custom metrics 
that are deemed relevant to the goals of some subsequent analyses, e.g., bug finding, on the system under analysis, which will utilize the configurations that will be generated as an output of CONFIZZ.

Unlike black-box sampling techniques such as Combinatorial Interaction Testing (CIT), 
CONFIZZ leverages the presence conditions to generate maximal configurations and 
to check the validity of the generated configurations. Additionally, 
it evaluates the importance of presence conditions under the guidance of some code metric.
CONFIZZ consists of three MaxSAT-based maximal configuration generation approaches, 
Maximal (Max), Maximal Iterative (MaxI), and Maximal Variants (MaxV), and 
three SAT-based configuration fuzzing approaches, Configuration Variable Fuzzing (CVF), 
Presence Condition Fuzzing (PCF), and Selective Presence Condition Fuzzing (SPCF).
The code metric allows the ranking of generated configurations for a specific goal. 
We use an adaptation of CIT by evaluating the generated configurations with respect to a given code metric using SAT solving.

We discuss a variety of code metrics and how we model their associations with the presence 
conditions in Section \ref{sec:codemetrics}. 
We briefly discuss the MaxSAT problem and its adaptation to the maximal configuration generation (Max) in Section \ref{sec:maximal}.
We present an approximation algorithm, MaxI, for 
generating the minimum number of configurations that cover all the presence conditions in Section 
\ref{sec:optimization}.
We explain the details of MaxV, CVF, PCF, and SPCF in Section \ref{sec:variantgen}.

\subsection{Code Metrics and Presence Conditions}
\label{sec:codemetrics}

\begin{figure}[th]
  \centering
 \begin{subfigure}[b]{0.2\textwidth}
\begin{footnotesize}  
\begin{Verbatim}[numbers=left,xleftmargin=3mm]
    #if C1
       free(p1);
    #endif

    #if C2
      #if C1 
        free(p2);
        x++; 
        printf(...);
      #else
        free(p3);
      #endif  
    #endif   
\end{Verbatim}
\end{footnotesize}
    \caption{}
    \label{fig:code1}
  \end{subfigure}
  \hspace{0.5cm}
 \begin{subfigure}[b]{0.2\textwidth}
 \begin{footnotesize} 
\begin{Verbatim}[numbers=left,xleftmargin=3mm]
    #if C4 || C5
       free(p1);
    #endif

    #if C5
       printf(...)
    #endif

    #if !C5 && C6
       free(p2);
    #endif   

    #if C5
       printf(...)
       p = malloc(...)
    #endif
\end{Verbatim}
\end{footnotesize}
    \caption{}
    \label{fig:code2}
  \end{subfigure}
  \caption{Example code with various presence conditions.}
  \label{fig:codesamples}
%   \vspace{-0.5cm}
\end{figure}

A presence condition is a boolean expression on the compile-time configuration variables and is used in preprocessor statements such as {\tt \#if} and {\tt \#ifdef} as the 
condition expression.  
Figure \ref{fig:codesamples} shows example code snippets 
with presence conditions.
We represent each instance of a presence condition with the set of code locations it controls.
For example in Figure \ref{fig:code1}, presence conditions $C1$ and $C1 \wedge C2$ 
control lines \{2\} and lines \{7,8,9\}, respectively.
So, we represent $C1$ at line 1 with \{2\} and $C1 \wedge C2$ at lines 5-6 with 
\{7,8,9\}. 
 
Since the same presence condition can appear in multiple code locations, e.g., $C5$
in Figure \ref{fig:code2}, we use the set of sets notation to represent the source code locations controlled by a presence condition. 
For example, $C1$ and $C1 \wedge C2$, are mapped to 
\{\{2\}\}, \{\{7,8,9\}\}, respectively. 
Also, $C5$ is mapped to 
\{\{6\},\{14,15\}\} to cover its instances at lines 
5 and 13. 

Once this mapping is generated for a software component, it is desirable to 
quantify the role of a presence condition with respect to code coverage for that component. 
One possible way to quantify a given presence condition or assign a weight to it 
is to use the sum of the sizes of individual sets, e.g., the number of covered source code lines that corresponds to it. We call this the {\em absolute} view of the code coverage.
The weights of $C1$ and $C1 \wedge C2$ according to the absolute view are 1 and 3,  
respectively. 
Another way to measure the code coverage for a presence condition is to use the size of the set, which we call the {\em Logical Code Unit (LCU)}  
view of source code coverage. In logical code unit view, the weights of both 
presence conditions $C1$ and $C1 \wedge C2$ are 1 despite the fact that 
$C1 \wedge C2$ controls more source code lines (3) than those controlled by $C1$ (1). 
So, both \{2\} and \{7,8,9\} represent single logical code units as they are controlled 
by a unique presence condition. And the LCU weight of $C5$ is 2 although its absolute weight is 3.

The logical unit view can guide the testing process to quantify the {\em configuration code coverage} (CCU)
by computing the following ratio: 
\begin{equation}
\textit{CCU}=\frac{\#\textit{covered logical code units}}{\textit{total \# logical code units}}
\end{equation}

A configuration can be represented as the set of presence condition instances it enables, i.e., evaluates to true.
So, the CCU of a configuration is defined as the ratio of the number of logical code units covered 
by the enabled presence conditions to the total number of logical code units in the system under test.
As an example, there are a total of 4 logical code units in Figure \ref{fig:code2}.
Configuration $C4=1,C5=0,C6=0$ covers one logical code unit (\{\{2\}\}) achieving a 25\% CCU and 
configuration $C4=0,C5=1,C6=0$ covers three logical code units (\{\{2\},\{6\},\{14,15\}\}) achieving a 75\% CCU.

\begin{table}[th!]
    \centering
    \begin{tabular}{c|c|c|c|c} \hline
       {\bf Presence}  & \multicolumn{2}{c|}{\bf LCU Cov.} & \multicolumn{2}{c}{\bf Free Calls}\\ \cline{2-5}
       {\bf Condition} &  {\bf Code Unit} &  {\bf Weight} & {\bf Code Unit} & {\bf Weight}\\ \hline
       $C1$ & \{\{2\}\} & 1 & \{\{2\}\} & 1\\
       $C2 \wedge C1$ & \{\{7,8,9\}\} & 1& \{\{7\}\}& 1\\
       $C2 \wedge \neg C1$ & \{\{11\}\} & 1& \{\{11\}\} & 1\\ \hline
       $C4 \vee C5$ & \{\{2\}\} & 1 & \{\{2\}\} & 1\\
       $C5$ & \{\{6\},\{14,15\}\} & 2 & $\emptyset$ & 0\\
       $\neg C5 \wedge C6$ & \{\{10\}\} & 1 & \{\{10\}\} &1\\ \hline
       %$C7$ & \{\{2\}\} & 1 & $\emptyset$ & 0\\
       %$C8$ & \{\{6,7\}\} & 1 & \{\{7\}\} & 1\\
       %$C9$ & \{\{11\}\} & 1 & \{\{11\}\} & 1\\ \hline
    \end{tabular}
    \caption{Presence conditions and their code metrics.}
    \label{tab:prescond}
    %\vspace{-1cm}
\end{table}

For general purpose testing the logical view of code coverage may be sufficient. However, 
when there is a specific analysis goal such as finding a specific type of bug or 
focusing on the specific part of the code or configuration variables \cite{El-SharkawyYS19}, 
it is necessary to consider custom code metrics.
As an example, assume that our goal is to detect double-free bugs where the same memory object 
is deallocated with two consecutive calls to a memory deallocation function on pointers 
storing the same memory address, e.g., {\tt free},
without any initialization of the pointer with a valid memory address in between. 
Then for the code in Figure \ref{fig:code2}, we can ignore logical code units that are 
controlled by presence condition $C5$, which control calls to {\tt printf}s, and focus on those controlled by $C4 \vee C5$ and 
$\neg C5 \wedge C6$ as the code they control includes calls to {\tt free}.
In this case, we can define a code metric relevant to the goal of detecting double-free bugs 
by assigning weights to presence conditions based on the existence or quantity of the calls to {\tt free} and filtering them based on whether the weights exceed a user tunable threshold.  
Table \ref{tab:prescond} shows the code units related to two code metrics, logical 
view of code coverage and existence of {\tt free} calls,
and the weights of presence conditions for each unique presence condition.
We can filter out the presence condition $C5$ for double-free detection that uses 
the existence of {\tt free} calls as the code metric as its weight is 0 and we would want to 
focus on the presence conditions with positive weights, e.g., $C4 \vee C5$ and 
$\neg C5 \wedge C6$.

\subsection{Generating a Maximal Configuration}
\label{sec:maximal}

While it may not be possible to satisfy all the presence conditions at once, it is desirable to 
find a maximal set of presence conditions that can be satisfied at once. We can generate 
such a maximal configuration by encoding the problem as a MaxSAT problem. 
In a MaxSAT problem, there are two types of constraints: the hard constraints that must be 
satisfied and the soft constraints that are associated with some weights. 
There is also an 
objective function that guides the solver to choose the soft constraints to satisfy the 
objective function. We encode the maximal configuration finding problem as a MaxSAT problem by 
assuming the hard constraints to be true, using the presence conditions as the soft constraints, 
and the objective being maximizing the sum of presence condition weights, which are determined 
based on the given code metric.
When we use the LCU metric, the weight is computed as the size of the set representation that we discussed in \ref{sec:codemetrics}. 
As an example, for the code snippet in Figure \ref{fig:code2}, the soft constraints are
$C4 \vee C5$, $C5$,  $\neg C5 \wedge C6$ with weights 1, 2, and 1, respectively.
A maximal solution includes lines 2,6,14, and 15, and can be found by setting $C5$ to true, yielding a total weight of 3.
If we use the number of calls to {\tt free}, then the soft constraints 
$C4 \vee C5$, $C5$,  $\neg C5 \wedge C6$ would be assigned the weights 1, 0, 1, and 
a maximal solution includes lines 2 and 10 in the compiled code and can be found by setting $C4$ and $C6$ to true and $C5$ to false.
We use the Optimization API of Z3  \cite{BP14} to generate maximal configurations customized for a given code metric and call this 
the {\em Maximal (Max)} approach.

\subsection{Generating a Minimal Set of Configurations}
\label{sec:optimization}

While code metrics can help an analyst to reduce the configuration search space 
in accordance with the analysis goal, the size of the reduced configuration space
may still be intractable. 
It is thus desirable to be able to generate a minimum set of goal relevant configurations 
and ensure 100\% coverage of the configurable code.
However, this optimization problem is challenging to solve for a real-world system with thousands of configuration variables.
We formulate the generation of minimal configuration set 
as a minimum set covering problem, which is NP-hard, and present 
an approximation algorithm that uses MaxSAT solving \cite{BP14}.

In the minimum set covering problem, there is a universe set of elements, $U$, 
and a set, $S$, such that $\forall s \in S. s \subseteq U$ and $U=\bigcup_{s \in S} s$. 
Each $s$ in $S$ has a cost $C(s)$, according to some cost function $C$.
The goal is to find a minimal subset, $S_{min} \subseteq S$ such that 
all the elements in the universe are covered, $U=\bigcup_{s \in S_{min}} s$, and
the cost of $Cost(S_{min})=\Sigma_{s \in S_{min}} C(s)$, is maximized. 
Since the minimum set covering problem is NP-hard, we focus on a polynomial-time greedy approximation 
algorithm \cite{SLAV97} to make it practical for testing real-world software. 
The idea is to work in an iterative way and in each iteration choose the largest number of 
elements from the universe to include in the minimal set.

\begin{algorithm}
\begin{footnotesize}
\begin{algorithmic}[1]
\State {\bf Maximal}($P$: Presence Conditions,$C$: Cost Function, $FM$: Feature Model): (Configuration, Unsatisfied Presence Conditions)
\State $HC \gets FM$
\State $SC \gets \emptyset$
\State $W \gets \lambda sc. 0$
\For{each $p \in P$}
  \If{$C$ is LCU}
     \State Set $w$ to the number of occurrences of $p$ in SUT
  \Else
    \State $w \gets \Sigma_{l \in Loc(p)} C(l)$
  \EndIf
  \If{$w > 0$}
     \State $SC \gets SC \cup \{p\}$
     \State $W \gets W[p \mapsto w]$
  \EndIf
\EndFor
\State {\bf return} MaxSAT(HC,SC,W)\\ 
\State {\bf MaximalIterative}($P$: Presence Conditions,$C$: Cost Function, $FM$: Feature Model): Set of Configurations
\State $MinConfigs \gets \emptyset$
\State $CP \gets P$
\While{$CP \not = \emptyset$}
  \State $(Model,UP) \gets Maximal(CP,C,FM)$
  \State $MinConfigs \gets MinConfigs \cup \{Model\}$
  \State $CP \gets UP$
\EndWhile
\State {\bf return} $MinConfigs$
\end{algorithmic}
\end{footnotesize}
\caption{The {\bf Maximal} algorithm that uses the MaxSAT computation to find a maximal configuration and the greedy {\bf MaximalIterative} algorithm that calls Maximal to compute an approximate solution to finding a minimum set of configurations using the code metric cost function $C$.}
\label{alg:minconfig}
\end{algorithm}

We model the problem of finding the minimal set of configurations for a 
configurable software component with a set of binary configuration variables\footnote{Our approach can be easily extended to non-binary configuration variables.}, $V$, 
as a minimum set covering problem, where 
the cost function $C$ is defined over some code metric(s), 
e.g., the number of {\tt free} calls, 
and the universe $U$ represents the set of all logical code units with non-zero cost.

Each $s \in S$ represents the code elements controlled by a unique set of configurations of the 
software, where each configuration is an element from the configuration space, $CS=2^{|V|}$. 
We will interpret each element $cs$ of $CS$ 
to represent a solution of the configuration space such that $v \in cs \leftrightarrow  v=1 \ \text{ in } cs$. Let $F(cs)$ represent the formula version of the 
configuration. For example, assuming that $V=\{C1,C2\}$, for the configurations $cs_0=\emptyset$, $cs_1=\{C_2\}$, $cs_2=\{C1\}$, and $cs_3=\{C1,C2\}$, the corresponding formula are defined as 
$F(cs_0)=\neg C1 \wedge \neg C2$, $F(cs_1)=\neg C1 \wedge C2$, 
$F(cs_2)=C1 \wedge \neg C2$, and $F(cs_3)=C1 \wedge C2$.

Let $P$ denote all the satisfiable presence conditions in the configurable software\footnote{Dead code locations that are controlled by unsatisfiable presence conditions are automatically eliminated from the problem space.}.
Let $Loc(p)$ denote the set of code locations controlled by a presence condition $p \in P$.
Let $Loc(cs)$ denote  
the union of code locations that are controlled by every presence condition $p \in P$ that has a non-zero cost and is satisfied 
by the configuration represented by $cs$, 
i.e., $Loc(cs)=\bigcup_{p \in P, p \wedge F(cs) \not = false, C(p) \not = 0} \{Loc(p)\}$.
Finally, $S$ in the minimal set cover problem is defined by $\bigcup_{cs \in CS} \{Loc(cs)\}$.
The cost function is 
used in controlling the solution space to be relevant to the goal of the analysis by filtering 
the code locations that have non-zero cost with respect to the code metric.
Formulating our approach as a greedy approach to the minimum set covering problem 
allows us to claim the suboptimality of our approach, which we will further elaborate on in Section \ref{sec:eval}.

\begin{algorithm}
\begin{footnotesize}
\begin{algorithmic}[1]
\State {\bf MaximalVariants}($P$: Presence Conditions, $MC$: Minimal Configurations): Set of Configurations
\State $Var \gets \emptyset$
\State $CPC: MC \mapsto 2^P$
\For{each $mc \in MC$}
  \State $CPC=CPC[mc \mapsto \{ p \ | \  p \in P \ \wedge mc \textit{ enables } p\}]$
\EndFor
\For{each $mc \in MC$}
   \State $Alt \gets \bigcup_{mc' \in MC, mc \not = mc'} CPC[mc']$
   \For{$p_1 \in CPC[mc]$}
       \For{$p_2 \in Alt$}
          \State $v \gets CPC[mc] \ \cup \ \{p_2\} \ \setminus \{p_1\}$
          \If{$SAT(\bigwedge_{p \in v} p)$}
             \State $Var \gets Var \ \cup \ \{v\}$
          \EndIf
       \EndFor
   \EndFor
\EndFor
\State {\bf return} $Var$
\end{algorithmic}
\end{footnotesize}
\caption{The {\bf MaximalVariants} algorithm that generates variants from the configurations returned by the MaximalIterative algorithm.} 
\label{alg:maxvar}
\end{algorithm}

As an example, for the code in Figure \ref{fig:code2}, 
finding the minimal set of configurations to achieve 100\% logical 
code coverage can be formulated as 
\begin{equation}
\begin{split}
U=\{\{2\}, \{6\}, \{10\}, \{14,15\}\}\\
V=\{C4,C5,C6\}\\ 
CS=\mathcal{P}(V)\\
P=\{C4 \ \vee \ C5, C5, \neg C5 \ \wedge \ C6\}\\
CS=\{\emptyset, \{C4\},\{C5\},\{C6\},\{C4,C5\},\\
\{C5,C6\},\{C4,C6\},
\{C4,C5,C6\}\}\\
S=\{s_i \ | \ 0 \leq i \leq 7\},
\end{split}
\end{equation}
where 
\begin{equation}
\begin{split}
s_0=Loc(\emptyset)=\{\emptyset\}\\
s_1=Loc(\{C4\})=\{\{2\}\}\\
s_2=Loc(\{C5\})=\{\{2\},\{6\},\{14,15\}\}\\
s_3=Loc(\{C6\})=\{\{10\}\}\\
s_4=Loc(\{C4,C5\})=\{\{2\},\{6\},\{14,15\}\}\\
s_5=Loc(\{C5,C6\})=\{\{2\},\{6\},\{14,15\}\}\\
s_6=Loc(\{C4,C6\})=\{\{2\},\{10\}\}\\
s_7=Loc(\{C4,C5,C6\})=\{\{2\},\{6\},\{14,15\}\}.
\end{split}
\end{equation}

Our greedy algorithm would first choose one of $s_2$,$s_4$,$s_5$, or $s_7$ as each has the largest size among other elements in $S$. 
Then it would choose $s_6$ to also include line 10 and fully cover $U$. So, a possible approximate minimal set of configurations would be $\{\{C5\},\{C4,C6\}\}$, where 
\{C5\} representing configuration $C4=C6=0,C5=1$ and 
\{C4,C6\} representing configuration $C4=C6=1,C5=0$. 
Assuming that the cost function is defined as the 
number of logical code units for a given subset,
the total cost of the solution is 4, as 
$C(\{\{C5\},\{C4,C6\}\})=C((C5 \wedge \neg C4 \wedge \neg C6) \vee (C4 \wedge C6 \wedge \neg C5))$=
$|\{\{2\},\{6\},\{10\},\{14,15\}\}|=4$, and, hence, achieves 
100\% CCU using two configurations out of a configuration space of size eight. 

If our goal was to generate a minimum set of configurations with the goal of finding 
double-free bugs, as the code metric we could assign a score of one to the logical code units that include some {\tt free} calls and 0 to those without any {\tt free} calls and 
for the code in Figure \ref{fig:code2}, 
\begin{equation}
\begin{split}
U=\{\{2\}, \{10\}\}\\ 
V=\{C4,C5,C6\}\\ 
CS=\mathcal{P}(V)\\
P=\{C4 \ \vee \ C5, C5, \neg C5 \ \wedge \ C6\}\\
S=\{s_i \ | \ 0 \leq i \leq 7\},
\end{split}
\end{equation}
where 
\begin{equation}
\begin{split}
s_0=Loc(\emptyset)=\{\emptyset\}\\
s_1=Loc(\{C4\})=\{\{2\}\}\\
s_2=Loc(\{C5\})=\{\{2\}\}\\
s_3=Loc(\{C6\})=\{\{10\}\}\\
s_4=Loc(\{C4,C5\})=\{\{2\}\}\\
s_5=Loc(\{C5,C6\})=\{\{2\}\}\\
s_6=Loc(\{C4,C6\})=\{\{2\},\{10\}\}\\
s_7=Loc(\{C4,C5,C6\})=\{\{2\}\}.
\end{split}
\end{equation} 

The greedy algorithm would run for only one iteration and choose $s_6$, and, hence, return $\{\{C4,C6\}\}$ as the 
minimal solution, achieving 100\% coverage of code that call {\tt free} with a single configuration in a configuration space of size eight.

Algorithm \ref{alg:minconfig} presents our algorithm at a high level. 
If the code metric is LCU, each unique presence condition is associated with a weight equal to the number of LCUs it controls (line 7). For other code metrics that come with a cost function, $C$, that maps every  source line to some value, the weight of each unique presence condition is 
computed as the sum of the cost function for each source line it controls (line 9). 
Associating each presence condition with a weight as explained above allows the MaxSAT solver to implicitly perform the 
encoding of the problem as the minimum cover set problem as we explain previously. 
Each call to the MaxSAT solver (line 16) in {\em Maximal} is an iteration of our adaptation of the greedy algorithm, which is implemented in {\em MaximalIterative}. 
Given the set of presence conditions, the algorithm computes the maximal solution among the 
current set of presence conditions $CP$ (line 22) and uses the unsatisfied presence conditions $UP$ as the set of presence conditions to be considered for the next iteration (line 24). 
In each iteration, the configuration that achieves the highest cost is returned by {\em Maximal}. The set of configurations that are returned by {\em MaximalIterative} is guaranteed to achieve 100\% code coverage and coverage of all the presence conditions with nonzero weight. 
However, due to the approximate nature of {\em MaximalIterative}, it may not necessarily be the minimum set of configurations that achieves 100\% code coverage. 

\subsection{Generating Maximal Variants}
\label{sec:variantgen}

The Maximal Iterative approach provides a partitioning of the configuration space that 
can be used as a starting point to generate variants. 
Algorithm \ref{alg:maxvar} shows the generation of variants of the maximal configurations 
that get generated by the {\em MaximalIterative} algorithm. Basically, for each configuration $mc$
in the input set $MC$, variants that have the same size, i.e., the number of presence conditions 
satisfied, are generated by replacing each presence condition in $mc$ with a presence condition from those enabled by other minimal configurations, $Alt$, in $MC$ (lines 10-11). 
If the presence conditions in the variant are consistent (line 8) then it is included in 
the variants set $Var$.

\begin{algorithm}
\begin{footnotesize}
\begin{algorithmic}[1]
\State {\bf ConfigurationFuzzing}($P$: Presence Conditions, $type: \{\textit{confvar}, \textit{prescond}\}$,  $C$: Cost Function, $SConf$: Seed Configurations, $cycles$: Integer, $M$: Num to be filtered)
\State $\textit{PriorityQueue} \ \textit{queue}, \textit{confqueue} \gets \emptyset$
\For{each seed $\textit{sc}$ in $\textit{SConf}$}
   \State $CheckAndAdd(\textit{queue},\textit{confqueue},\textit{sc})$
\EndFor     
\If{$type$ is confvar} Set $Max$ to number of configuration vars
\Else \ Set $Max$ to number of unique presence conditions
\EndIf
\State $\textit{cycle} \gets 1$
\While{$\textit{cycle} \leq cycles$ and $\textit{queue}$ not empty}
   \State $\textit{cur} \gets \textit{queue}.\textit{removeMax}()$
    \For{$i$ in \{1,2,3,...,$Max$\}}
    \State Flip consecutive $i$ bit(s) in $\textit{cur}$ and store in $\textit{cur'}$
     \State $CheckAndAdd(\textit{queue},\textit{confqueue},\textit{cur'})$
  \EndFor
   \For{$i$ in \{1,2,3,...,$Max$\}}
     \State Flip $i$ random bit(s)  in $\textit{cur}$ and store in $\textit{cur'}$
      \State $CheckAndAdd(\textit{queue},\textit{confqueue},\textit{cur}')$
   \EndFor   
   \State $cycle++$
\EndWhile
\State {\bf return} Top $M$ configurations in $\textit{confqueue}$
\State
\State {\bf Score}($\textit{conf}$: String, $C$: Cost Function): $\mathcal{N}$
   \State Let $pcs$ denote the presence conditions enabled by $conf$
   \If{$C=LCU$} 
     {\bf return}  $|\textit{pcs}|$
   \Else 
        \ {\bf return} $\Sigma_{p \in \textit{pcs}, l \in Loc(p)} C(l)$
   \EndIf
\State   
\end{algorithmic}
\begin{algorithmic}[1]
\State {\bf CheckAndAdd($\textit{fq}$, $\textit{confq}$: PriorityQueue , $\textit{conf}$: String)}
   \State Let $pcs$ denote the presence conditions enabled by $conf$
\State $\textit{pcf} \gets \bigwedge_{\textit{pc} \in \textit{pcs}} \textit{pc}$
\If{SAT($\textit{pcf}$)}
  \State $\textit{model} \gets getModel(pcf)$
  \State $score \gets \textit{Score}(\textit{conf},C)$
  \State Add $(\textit{conf},\textit{score})$ to $\textit{fq}$
   \State Add $(\textit{model},\textit{score})$ to $\textit{confq}$
\EndIf
\end{algorithmic}
\end{footnotesize}
\caption{The {\bf Configuration Fuzzing} algorithm that implements the Configuration Variable Fuzzing approach (CVF) when the type is \textit{confvar} and implements the Presence Condition Fuzzing (PCF)
approach when the type is \textit{prescond}.}
\label{alg:fuzzing}
\end{algorithm}

\subsection{Configuration Fuzzing}
\label{sec:conffuzz}

An alternative approach to generating maximal configurations is configuration fuzzing. 
Given a binary encoding of the configuration space, deterministic and non-deterministic 
mutations, i.e., bit flips, are applied to generate candidate configurations. 
Algorithm \ref{alg:fuzzing} starts with a set of seed configurations $SConf$ that get added to a priority 
queue based on their score according to the given code metric $C$.
Each candidate configuration maps to a set of presence conditions, and the 
feasibility of the configuration is checked using a SAT solver. 
For those that are feasible, the score is computed based on $C$ and added 
to the priority queue. This is repeated for a given number of cycles, and the algorithm returns 
the top ranking configurations according to $C$.

Although mutating the configuration variables 
is a natural way for searching the configuration space, the configuration space 
defined by the configuration variables may be 
much larger than the one defined by the presence conditions. 
For instance, in BusyBox version 1.36, there are 1077 configuration variables defined while 
the maximum number of unique number of presence conditions in our benchmark set is 156 (happens to be in the {\tt ls} component). 
Therefore, we consider both type of fuzzing in this paper.

Algorithm \ref{alg:fuzzing} is parameterized by the type of fuzzing. If the type is {\tt confvar} 
then it implements the Condition Variable Fuzzing approach, where each boolean configuration variable has a fixed position in the configuration string that gets mutated.
If the type of {\tt prescond} then it implements the Presence Condition Fuzzing approach, where 
each unique presence condition has a fixed position in the configuration string that gets mutated.
A variation of Presence Condition Fuzzing is Selective Presence Condition Fuzzing (SPCF), where the presence conditions that have nonzero scores for the code metric get mutated. 

As an example, for the code in Figure \ref{fig:code2}, performing single bit flips 
for Condition Variable Fuzzing 
on a seed configuration $(C4=0,C5=0,C6=0)$ with an LCU score of 0 would yield the candidates 
$(C4=1,C5=0,C6=0)$, $(C4=0,C5=1,C6=0)$, and $(C4=0,C5=0,C6=1)$ with LCU scores of 1, 3, and 1, respectively, which get added to the priority queue and giving  $(C4=0,C5=1,C6=0)$ a higher chance for the application of additional mutations.
Similarly, performing single bit flips for Presence Condition Fuzzing 
on a seed configuration $(C4 \vee C5=false, C5=false,\neg C5 \wedge C6=false)$ 
with an LCU score of 0 would yield feasible candidates 
$(C4 \vee C5=true, C5=false,\neg C5 \wedge C6=false)$,
and $(C4 \vee C5=false, C5=false,\neg C5 \wedge C6=true)$ 
 with LCU scores of 1 and 1, respectively, 
and an infeasible configuration
$(C4 \vee C5=false, C5=true,\neg C5 \wedge C6=false)$.

\begin{table}[th!]
\centering
\begin{footnotesize}
\begin{tabular}{c|c|c} \toprule 
    {\bf Component} & \multicolumn{2}{c}{\bf \% PC Cov.} \\ \cline{2-3} 
     & {\bf DefConfig} & {\bf Maximal Config} \\ \hline
     awk & 61.29 & 70.97 \\
     cmp & 63.37 & 71.29\\
     diff & 62.63 & 73.73\\
     ed & 61.54 & 71.43\\
     patch\_bbox & 61.86 &  73.20 \\
     patch\_toybox& 61.54 & 71.43\\
     patch & 61.86 &  73.20 \\
     sed & 61.86 & 71.13 \\    \bottomrule 
\end{tabular}
\end{footnotesize}
\caption{Comparison of defconfig and the MaxSAT-generated configuration in terms of  presence condition (PC) coverage for the components in BusyBox {\em editors} subsystem.}
\label{tab:defvsmaxsateditors}
\end{table}

\begin{table}[th!]
\centering
\begin{footnotesize}
\begin{tabular}{c|r|r} \toprule 
    {\bf Component} & \multicolumn{2}{c}{\bf \% PC Cov.} \\ \cline{2-3} 
     & {\bf DefConfig} & {\bf Maximal Config} \\ \hline
%      basename & 61.80 & 71.91  \\ 
      cat &  63.00& 74.00  \\ 
%      chgrp &  61.80 & 71.91  \\ 
%      chmod &  61.54 & 71.43  \\ 
      chown &  61.86 & 73.20  \\ 
%      chroot & 61.80 & 71.91  \\ 
%      cksum &  62.37 & 72.04  \\ 
%      comm & 61.80  & 71.91  \\ 
      cp &  59.62 & 70.19  \\ 
%      cut & 62.10  & 69.47  \\ 
%      date & 60.19 & 68.93  \\ 
%      dd & 64.70 & 68.07  \\ 
      df & 53.57 & 62.50 \\ 
%      dirname &  61.80 & 71.91  \\ 
%      dos2unix &  61.80 & 71.91  \\ 
%      du & 61.05 & 70.53   \\ 
      echo &  61.29 & 70.97  \\ 
%      env &  61.54 & 71.43 \\ 
      expand & 54.31 & 67.24  \\ 
%      expr & 61.29 & 70.97  \\ 
%      factor & 61.80 & 71.91  \\
%      false & 61.80 & 71.91  \\
%      fold & 53.57 & 66.96  \\ 
      head & 62.74 & 68.63  \\
%      hostid & 61.80 & 71.91  \\
%      id &  60.00 & 69.47  \\ 
%      install & 60.40 & 72.28  \\
%      link & 61.53 & 71.43 \\ 
%      ln & 61.53 & 71.43  \\
%      logname & 61.80 & 71.91  \\
      ls & 58.33 & 62.18  \\
%      md5\_sha1\_sum&  61.05 & 70.52  \\
    \bottomrule 
\end{tabular}
\end{footnotesize}
\caption{Comparison of defconfig and MaxSAT generated config in terms of presence condition (PC) coverage for the components in BusyBox {\em coreutils} subsystem.}
\label{tab:defvsmaxsatcoreutils}
\end{table}

\section{Evaluation}
\label{sec:eval}

We have implemented CONFIZZ on top of SuperC \cite{GazzilloG12} and using Z3's Optimization library. SuperC uses Binary Decision Diagrams (BDDs) to represent the presence conditions. 
We converted BDDs to Z3 expressions to use the SAT solver or the MaxSAT solver from Z3.
We used BusyBox version 1.36.0 to answer the research questions (RQs) presented  in Section \ref{sec:intro}. We used eight components from the {\em coreutils} subsystem and eight components from the {\em editors} subsystem.
We present the results of our evaluation and elaborate on each RQ in the following subsections.

\subsection{RQ1 on the Default Configuration}
\label{sec:rq1}
To answer {\bf RQ1: What percentage of the presence conditions in each component of the BusyBox benchmarks are covered by the default configuration?}, 
 we checked the number of presence conditions that are enabled by the default configuration using a SAT solver. 
 Tables \ref{tab:defvsmaxsateditors} and \ref{tab:defvsmaxsatcoreutils} show that the default configurations cover on average 61.99\% 
 and 59.34\% of the presence conditions for the editors and the coreutils components, respectively, while the maximal configurations based on the Z3's MaxSAT solver cover 
 72.05\% and 68.61\% of the presence conditions for the editors and the coreutils components, respectively. This indicates that 
 the default configuration 
  can be improved upon to achieve 
 better coverage in individual components. 

\begin{table}[th!]
    \centering
    \begin{footnotesize}
    \begin{tabular}{c|r|r|r|r|c} \toprule
      {\bf Component}  & {\bf \#PC} & \multicolumn{2}{c|}{\bf Maximal PC Cov} & \multicolumn{2}{c}{\bf Maximal Iterative } \\
      & & {\bf \# PC} & {\bf \% PC} & {\bf \# Config } & {\bf \# PCs} \\ \hline
      awk & 93 & 66 & 70.97 & 3 & 66, 26, 1  \\ 
      cmp & 101 & 72 & 71.29 & 2 & 72, 29 \\ 
      diff & 99 & 73 & 73.73 & 3 & 73, 25, 1 \\ 
      ed & 91 & 65 & 71.43 & 3 & 65, 25, 1 \\ 
      patch\_bbox & 97 & 71 & 73.20 & 3 & 71, 23, 3 \\ 
      patch\_toybox & 91 & 65  & 71.43 & 3 & 65, 25, 1 \\
      patch & 97 & 71 &  73.20 & 3 & 71, 23, 3 \\
      sed & 97 & 69 & 71.13 & 3 & 69, 27, 1 \\
      \bottomrule 
    \end{tabular} 
    \end{footnotesize}
    \caption{The presence condition (PC) coverage of MaxSAT generated configuration and the number of configurations to cover all the presence conditions using MaxSAT-Based Approximate (MBA) and their sizes for each component of BusyBox {\em editors} subsystem.}
    \label{tab:maxsatandapproximateeditors}
\end{table}

\begin{table}[th!]
    \centering
        \begin{footnotesize}
    \begin{tabular}{c|r|r|r|r|c} \toprule
      {\bf Component}  & {\bf \#PC} & \multicolumn{2}{c|}{\bf Maximal PC Cov} & \multicolumn{2}{c}{\bf Maximal Iterative } \\
      & & {\bf \# PC} & {\bf \% PC} & {\bf \# Config } & {\bf \# PCs} \\ \hline
%      basename & 89 & 64 & 71.91 & 2 & 64, 25 \\ 
      cat & 100 & 74 & 74.00 & 3 & 74, 25, 1 \\ 
%      chgrp & 89 & 64 & 71.91 & 2 & 64, 25 \\ 
%      chmod & 91 & 65 & 71.43 & 3 & 65, 25, 1 \\ 
      chown & 97 & 71 & 73.20 & 3 & 71, 24, 2 \\ 
%      chroot & 89 & 64 & 71.91 & 2  & 64, 25 \\ 
%      cksum & 93 & 67 & 72.04  & 2 & 67, 26 \\ 
%      comm & 89 & 64 & 71.91 &2  & 64, 25 \\ 
      cp & 104 & 73 & 70.19 & 5 & 73, 25, 3, 2, 1 \\ 
%      cut & 95 & 66 & 69.47 & 3 & 66, 28, 1 \\ 
%      date & 103 & 71 & 68.93 &3  & 71, 30, 2 \\ 
%      dd & 119 & 81 & 68.07 & 3 & 81, 34, 4 \\ 
      df & 112 & 70 & 62.5 & 3 & 70, 39, 3\\ 
%      dirname & 89 & 64 & 71.91 & 2 & 64, 25 \\ 
%      dos2unix & 89 & 64 & 71.91 & 2 & 64, 25 \\ 
%      du & 95 & 67 & 70.53 & 3 & 67, 27, 1  \\ 
      echo & 93 & 66 & 70.97 & 2& 66, 27 \\ 
%      env & 91 & 65 & 71.43 & 2 & 65, 26 \\ 
      expand & 116 & 78 & 67.24 & 3& 78, 35, 3 \\ 
%      expr & 93 & 66 & 70.97 & 3 & 66, 26, 1 \\ 
%      factor & 89 & 64 & 71.91 & 2 & 64, 25 \\
%      false & 89 & 64 & 71.91 & 2 & 64, 25 \\
%      fold & 112 & 75 & 66.96 & 4 & 75, 34, 2, 1 \\ 
      head & 102 & 70 & 68.63 & 3 & 70, 24, 8 \\
%      hostid & 89 & 64 & 71.91 & 2 & 64, 25 \\
%      id &  95 & 66 & 69.47 & 3 & 66, 27, 2 \\ 
%      install & 101 & 73 & 72.28 & 3 & 73, 24, 4  \\
%      link & 91 & 65 & 71.43 & 3 & 65, 25, 1 \\ 
%      ln & 91 & 65 & 71.43 & 3 & 65, 25, 1 \\
%      logname & 89 & 64 & 71.91 & 2 & 64, 25 \\
      ls & 156 & 97 & 62.18 & 4 & 97, 53, 5, 1 \\
%      md5\_sha1\_sum& 95 &  67 & 70.52 & 4 & 67, 26, 1, 1 \\
      \bottomrule 
    \end{tabular} 
        \end{footnotesize}
    \caption{The presence condition (PC) coverage of MaxSAT generated configuration and the number of configurations to cover all the presence conditions using MaxSAT-Based Approximate (MBA) and their sizes for each component of BusyBox {\em coreutils} subsystem.}
    \label{tab:maxsatandapproximatecoreutils}
\end{table}

\begin{table*}[th!]
    \centering
    \begin{footnotesize}
    \begin{tabular}{c|c|c|c|c|c|c|c|c|c|c|c} \toprule
        {\bf Method} &  \multicolumn{5}{c|}{\bf 1L-Free (Score,Step)} & \multicolumn{5}{c|}{\bf 3L-Free (Score,Step)} &  \\  \cline{2-11}
                      & cp & df & expand & head & ls & cp & df & expand & head & ls & {\bf \# Best}\\ \hline
        {\bf Maximal} & (1,1) & (2,1) & (2,1) & (4,1) & (3,1) & (1,1) & (2,1) & (2,1) & (4,1) & (5,1) & 10 \\   
        {\bf Maximal It.}  & (1,1) & (2,1) & (2,1) & (4,1) & (3,1) & (1,1) & (2,1) & (2,1) & (4,1) & (5,1) & 10\\       
        {\bf Maximal Var.} & (1,1) & (2,1) & (2,1) & (4,1) & (3,1) & (1,1) & (2,1) & (2,1) & (4,1) & (5,1) & 10\\        
        {\bf PC Fuzz} & (1,49) & (2,67) & (2,1424) & (4,43) & (3,14) 
                      & (1,49) & (2,67) & (2,1253) & (4,43) & (5,3301) & 0 \\        
        {\bf Sel. PC Fuzz} & (1,1) & (2,1) & (2,3) & (4,1) & (3,1) 
                           & (1,1) & (2,1) & (2,3) & (4,1) & (5,13) & 7\\   
        {\bf CIT 2-way} & (1,24) & (2,24) & (2,24) & (4,24) & (3,24) 
                        & (1,24) & (2,24) & (2,24) & (4,24) & (5,24) & 0\\      
        {\bf CIT 3-way} & (1,92) & (2,92) & (2,92) & (4,92) & (3,92) 
                        & (1,92) & (2,92) & (2,92) & (4,92) & (5,92) & 0 \\      
        {\bf DefConfig} & (1,1) & (2,1) & (2,1) & (4,1) & (3,1) & (1,1) & (2,1) & (2,1) &  (4,1) & (4,1) & 9\\      
        \bottomrule
    \end{tabular}
        \end{footnotesize}
    \caption{Comparison of various configuration generation methods in terms of the maximum scores and the step they are reported for the 1L-Free and 3L-Free metrics. Number of cycles for fuzzing=5. \# Best denotes the number of maximum score achieved with the minimal \# of steps.}
    \label{tab:free_methodcmp_coreutils}
\end{table*}

\subsection{RQ2 on Full Coverage}
\label{sec:rq2}
To answer {\bf RQ2: How many configurations are needed for each component in the BusyBox benchmark set to cover all the presence conditions in the component?}, 
we used the Maximal Iterative approach. Tables \ref{tab:maxsatandapproximateeditors}
and \ref{tab:maxsatandapproximatecoreutils} show that 100\% presence coverage can be 
achieved using only three different configurations for most of the components. For {\tt cp}, 
five configurations suffice, and for {\tt ls} four configurations suffice, to cover all the presence conditions.

\begin{table}[th!]
    \begin{tabular}{c|c|r|r|r|r} \toprule
    {\bf Comp} & {\bf Config} & \multicolumn{2}{c|}{\bf Code Metric Score} & \multicolumn{2}{c}{\bf Fuzzing Results} \\ 
    & & 1L-Free & 3L-Free & \# Paths & \# Crashes \\ \hline
    awk & defconfig & 0 & 1 & 2403 & 262 \\
        & Maximal & 0 & 1 & 2337 & 236 \\
    diff & defconfig & 6 & 5 & 183 & 0 \\ 
         & Maximal & 6 & 5 & 199 & 0 \\
         & Maximal It. & 0 & 0 & 176 & 0 \\
    sed  & defconfig & 0 & 0 & 1805 & 0 \\ 
         & Maximal & 5 & 6 & 1747 & 106 \\
         & Maximal It. & 0 & 0 & 1708 & 0 \\
         \bottomrule
    \end{tabular}
    \caption{Results of fuzzing Busybox components for 8 hours using defconfig and configurations generated by the Maximal and Maximal Iterative approaches.}
    \label{tab:busyboxfuzzing}
\end{table}

\begin{table}[th!]
    \centering
    \begin{footnotesize}
    \begin{tabular}{c|c|c|c|c|c|c} \toprule
        {\bf Method} &  \multicolumn{2}{c|}{\bf 1L-Free (Score,Step)} & \multicolumn{3}{c|}{\bf 3L-Free (Score,Step)} & \\  \cline{2-6}
                      & {\bf diff} & {\bf sed} & {\bf awk} & {\bf diff} & {\bf sed} & {\bf \#Best}\\ \hline
        {\bf Max.}          & (6,1) &  (5,1) & (1,1) & (5,1) & (6,1) & 5 \\
        {\bf MaxI}      & (6,1) &  (5,1) & (1,1) & (5,1) & (6,1) & 5\\
        {\bf MaxV}     & (6,1) &  (5,373) & (1,1) & (5,1) & (6,373) & 3\\
        {\bf PCF}         & (6,57) & (5,25) & (1,58) & (5,57) & (6,25) & 0 \\ 
        {\bf SPCF}    & (6,1) &  (5,1) & (1,1) & (5,1) & (6,1) & 5\\         
        {\bf CIT2}       & (6,24) &  (5,24) & (1,24) & (5,24) & (6,24) & 0\\ 
        {\bf CIT3}       & (6,92) &  (5,92) & (1,92) & (1,92) & (6,92) & 0 \\ 
        {\bf DC}       & (6,1) & (0,1) & (1,1) & (5,1) & (0,1) & 3 \\
        \bottomrule
    \end{tabular}
        \end{footnotesize}
    \caption{Comparison of various configuration generation methods in terms of the maximum scores and the step they are reported for the 1L-Free and 3L-Free metrics. Number of cycles for fuzzing=5. \# Best denotes the number of maximum score achieved with the minimal \# of steps. DC stands for Default Configutaion.}
    \label{tab:free_methodcmp_editors}
\end{table}

\subsection{RQ3 on Comparison of Approaches}
To answer {\bf RQ3: How do the sampling, MaxSAT-based, and fuzzing approaches compare in terms of their maximum coverage scores and running times?}, 
we first analyzed the fuzzing approaches, Configuration Variable Fuzzing (CVF) and Presence Condition Fuzzing (PCF) to determine the number of cycles that yields the maximum scores for the logical code unit (LCU) coverage for a varying number of cycles, repeating each fuzzing session for five times. 
As shown in Figures \ref{fig:editors_cyclesmax} and \ref{fig:coreutils_cyclesmax}, for PCF the average maximum LCU coverage is achieved when the number of cycles is 25 for most of the components. We found that CVF is much slower (at least 30 times)  compared to  PCF and ran out of memory on some occasions, so we decided to run CVF for 5 cycles in the experiments. This is expected considering the difference in the sizes of the search spaces for the two approaches, $2^{1077}$ versus $2^{156}$ (max), even though each type of fuzzing tries to cover only a part of the search space.
We didn't include Maximal Iterative as the maximum score it achieves is always the same 
as Maximal.

To include a black-box sampling approach in the comparisons, we chose Combinatorial Interaction Testing (CIT), which is an effective approach for detecting feature interactions for highly configurable software, with reasonable testing efforts. Kuhn, et al. \cite{Kuhn2010} showed that as high as 90\% of feature interaction failures in various domains could be detected by 2-way (pairwise) and 3-way combinations of features, reaching 100\% with 4 to 6-way interactions.  For comparison with our proposed CONFIZZ approach, we used the ACTS tool,  a widely used CIT tool, to generate 2-way (CIT2) and 3-way (CIT3) combinations of all Boolean configurable features for Busybox-1.36.0, using In-parameter-order-general-doubling-construction (IPOG-D) algorithm \cite{lei2008}. We used the configurations generated by CIT2 and CIT3 to compute their LCU scores by checking their consistency with the presence conditions using a SAT solver.

\begin{figure*}[th!]
\centering
\begin{footnotesize}
\begin{tikzpicture}
\begin{axis}[ybar,
        bar width=8pt,
        width=18cm,
        height=6cm,
        ymin=0,
        ymax=75,        
        ylabel={Ave. Max. Score},
        xtick=data,
        xticklabels = {
patch\_box,
patch\_toybox,
patch,
sed
        },
        xticklabel style={yshift=-2ex},
        major x tick style = {opacity=0},
        minor x tick num = 1,
        minor tick length=0.2ex,
        every node near coord/.append style={
                anchor=east,
                rotate=90
        }
        ]
\addplot[draw=black,fill=black!10, nodes near coords=CVF] table[x index=0,y index=1] \editorscompmethb;
\addplot[draw=black,fill=black!15, nodes near coords=DefConf] table[x index=0,y index=2] \editorscompmethb;
\addplot[draw=black,fill=black!20, nodes near coords=CIT2] table[x index=0,y index=3] \editorscompmethb;
\addplot[draw=black,fill=black!25, nodes near coords=CIT3] table[x index=0,y index=4] \editorscompmethb;
\addplot[draw=black,fill=black!30, nodes near coords=Max] table[x index=0,y index=5] \editorscompmethb;
%\addplot[draw=black,fill=black!20, nodes near coords=MSI] table[x index=0,y index=5] \editorscompmeth;
\addplot[draw=black,fill=black!35, nodes near coords=MaxV] table[x index=0,y index=6] \editorscompmethb;
\addplot[draw=black,fill=black!40, nodes near coords=PCF] table[x index=0,y index=7] \editorscompmethb;
\end{axis}
 \end{tikzpicture}
 \end{footnotesize}
\caption{Comparison of methods w.r.t. LCU max score for patch\_bbox-sed. PCF (cycles=25). CVF (cycles=5).}
\label{fig:editors2_cyclesmax}
\end{figure*}
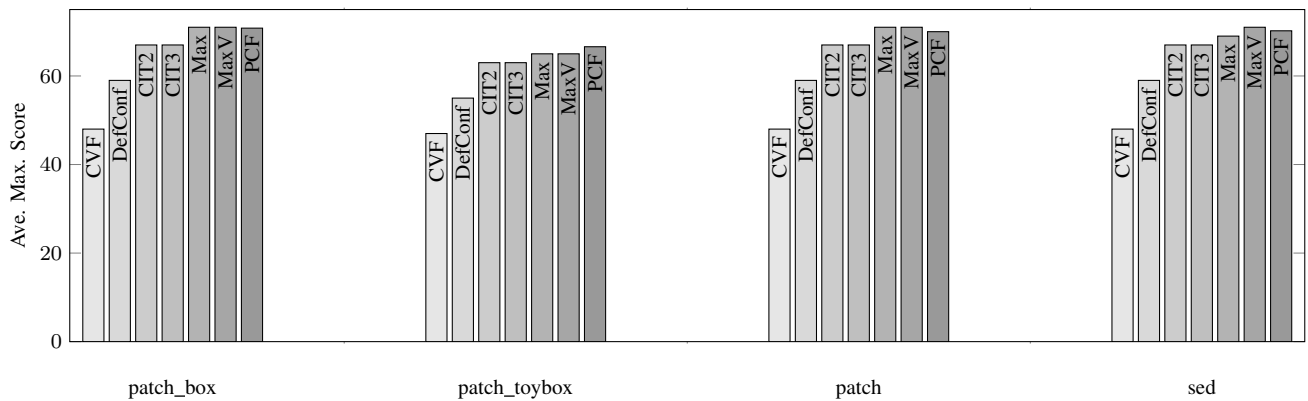

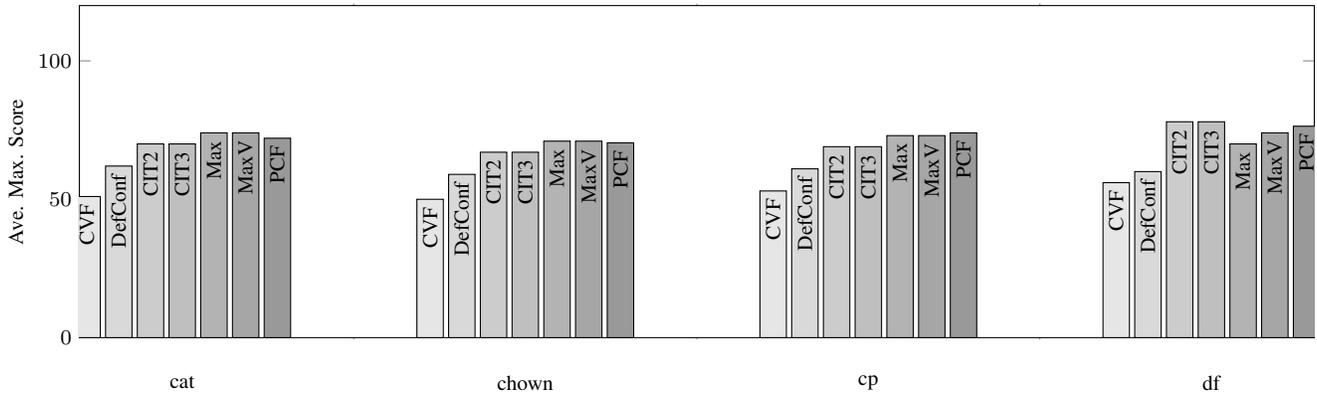
\begin{figure*}[th!]
\centering
\begin{footnotesize}
\begin{tikzpicture}
\begin{axis}[ybar,
        width=18cm,
        height=6cm,
        ymin=0,
        ymax=120,        
        ylabel={Ave. Max. Score},
        xtick=data,
        xticklabels = {
cat,        
chown,
cp,
df
        },
        xticklabel style={yshift=-2ex},
        major x tick style = {opacity=0},
        minor x tick num = 1,
        minor tick length=0.2ex,
        every node near coord/.append style={
                anchor=east,
                rotate=90
        }
        ]
\addplot[draw=black,fill=black!10, nodes near coords=CVF] table[x index=0,y index=1] \coreutilscompmetha;        
\addplot[draw=black,fill=black!15, nodes near coords=DefConf] table[x index=0,y index=2] \coreutilscompmetha;
\addplot[draw=black,fill=black!20, nodes near coords=CIT2] table[x index=0,y index=3] \coreutilscompmetha;
\addplot[draw=black,fill=black!25, nodes near coords=CIT3] table[x index=0,y index=4] \coreutilscompmetha;
\addplot[draw=black,fill=black!30, nodes near coords=Max] table[x index=0,y index=5] \coreutilscompmetha;
%\addplot[draw=black,fill=black!20, nodes near coords=MSI] table[x index=0,y index=5] \editorscompmeth;
\addplot[draw=black,fill=black!35, nodes near coords=MaxV] table[x index=0,y index=6] \coreutilscompmetha;
\addplot[draw=black,fill=black!40, nodes near coords=PCF] table[x index=0,y index=7] \coreutilscompmetha;
\end{axis}
 \end{tikzpicture}
 \end{footnotesize}
\caption{Comparison of methods w.r.t. LCU max score for cat-df. PCF (cycles=25). CVF (cycles=5).}
\label{fig:coreutils1_cyclesmax}
\end{figure*}

\begin{figure*}[th!]
\centering
\begin{footnotesize}
\begin{tikzpicture}
\begin{axis}[ybar,
        width=18cm,
        height=6cm,
        ymin=0,
        ymax=120,        
        ylabel={Ave. Max. Score},
        xtick=data,
        xticklabels = {
echo,        
expand,
head,
ls
        },
        xticklabel style={yshift=-2ex},
        major x tick style = {opacity=0},
        minor x tick num = 1,
        minor tick length=0.2ex,
        every node near coord/.append style={
                anchor=east,
                rotate=90
        }
        ]
\addplot[draw=black,fill=black!10, nodes near coords=CVF] table[x index=0,y index=1] \coreutilscompmethb;        
\addplot[draw=black,fill=black!15, nodes near coords=DefConf] table[x index=0,y index=2] \coreutilscompmethb;
\addplot[draw=black,fill=black!20, nodes near coords=CIT2] table[x index=0,y index=3] \coreutilscompmethb;
\addplot[draw=black,fill=black!25, nodes near coords=CIT3] table[x index=0,y index=4] \coreutilscompmethb;
\addplot[draw=black,fill=black!30, nodes near coords=Max] table[x index=0,y index=5] \coreutilscompmethb;
%\addplot[draw=black,fill=black!20, nodes near coords=MSI] table[x index=0,y index=5] \editorscompmeth;
\addplot[draw=black,fill=black!35, nodes near coords=MaxV] table[x index=0,y index=6] \coreutilscompmethb;
\addplot[draw=black,fill=black!40, nodes near coords=PCF] table[x index=0,y index=7] \coreutilscompmethb;
\end{axis}
 \end{tikzpicture}
 \end{footnotesize}
\caption{Comparison of methods w.r.t. LCU max score for echo-ls. PCF (cycles=25). CVF (cycles=5).}
\label{fig:coreutils2_cyclesmax}
\end{figure*}
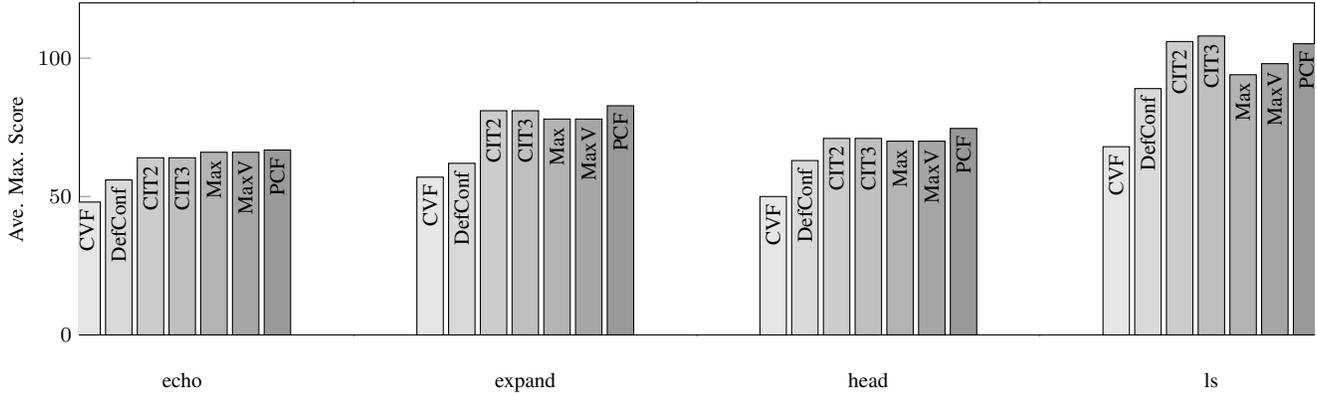

As Figures \ref{fig:editors1_cyclesmax} and \ref{fig:editors2_cyclesmax} show for the {\em editors} components, Maximal (Max) and Maximal Variants (MaxV) approaches generated the highest LCU coverage scores. PCF scores are also very close to those of Max and MaxV. While CIT2 and CIT3 show similar performance they do worse than Max, MaxV, and PCF while outperforming the default configuration (DefConf). 
For the {\em coreutils} components, we see a different trend as shown in Figures \ref{fig:coreutils1_cyclesmax} and \ref{fig:coreutils2_cyclesmax}. PCF achieves the 
highest coverage among the contenders with CIT2 and CIT3. Although in theory Max and MaxV should always have the maximum scores, 
Z3's Optimization API that we used may be providing suboptimal results in some cases as has been reported previously\footnote{See \url{https://github.com/Z3Prover/z3/issues/433}}.

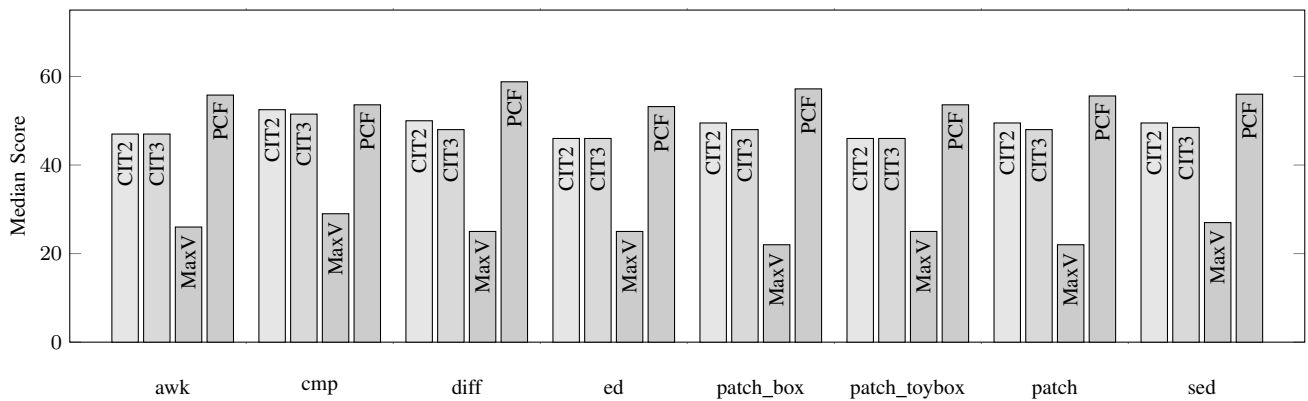
\begin{figure*}[th!]
\centering
\begin{footnotesize}
\begin{tikzpicture}
\begin{axis}[ybar,
        width=18cm,
        height=6cm,
        ymin=0,
        ymax=75,        
        ylabel={Median Score},
        xtick=data,
        xticklabels = {
awk,        
cmp,
diff,
ed,
patch\_box,
patch\_toybox,
patch,
sed
        },
        xticklabel style={yshift=-2ex},
        major x tick style = {opacity=0},
        minor x tick num = 1,
        minor tick length=0.2ex,
        every node near coord/.append style={
                anchor=east,
                rotate=90
        }
        ]
\addplot[draw=black,fill=black!10, nodes near coords=CIT2] table[x index=0,y index=1] \editorsmedian;        
\addplot[draw=black,fill=black!15, nodes near coords=CIT3] table[x index=0,y index=2] \editorsmedian;
\addplot[draw=black,fill=black!20, nodes near coords=MaxV] table[x index=0,y index=3] \editorsmedian;
\addplot[draw=black,fill=black!20, nodes near coords=PCF] table[x index=0,y index=4] \editorsmedian;
\end{axis}
 \end{tikzpicture}
 \end{footnotesize}
\caption{Comparison of CIT2, CIT3, MaxV, and PCF in terms of the median LCU score for awk-sed.}
\label{fig:editorsmedian}
\end{figure*}

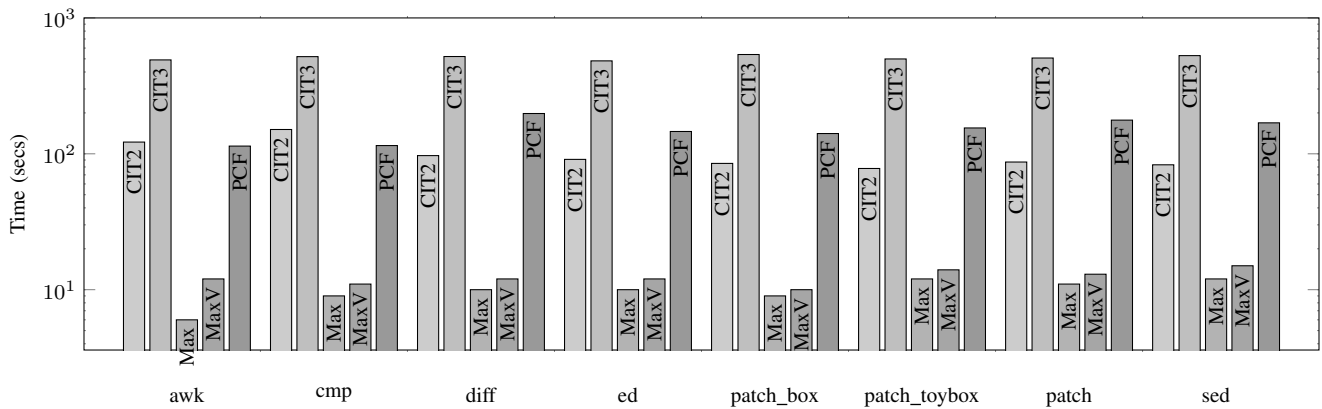
\begin{figure*}[th!]
\centering
\begin{footnotesize}
\begin{tikzpicture}
\begin{axis}[ybar,
        bar width=8pt,
        width=18cm,
        height=6cm,
        ymin=0,
        ymax=1000,        
        ylabel={Time (secs)},
        ymode=log,
        xtick=data,
        xticklabels = {
awk,        
cmp,
diff,
ed,
patch\_box,
patch\_toybox,
patch,
sed
        },
        xticklabel style={yshift=-2ex},
        major x tick style = {opacity=0},
        minor x tick num = 1,
        minor tick length=0.2ex,
        every node near coord/.append style={
                anchor=east,
                rotate=90
        }
        ]
%\addplot[draw=black,fill=black!10, nodes near coords=CVF] table[x index=0,y index=1] \editorscompmethtime1;        
%\addplot[draw=black,fill=black!15, nodes near coords=DC] table[x index=0,y index=2] \editorscompmethtime1;
\addplot[draw=black,fill=black!20, nodes near coords=CIT2] table[x index=0,y index=1] \editorscompmethtime;
\addplot[draw=black,fill=black!25, nodes near coords=CIT3] table[x index=0,y index=2] \editorscompmethtime;
\addplot[draw=black,fill=black!30, nodes near coords=Max] table[x index=0,y index=3] \editorscompmethtime;
%\addplot[draw=black,fill=black!20, nodes near coords=MSI] table[x index=0,y index=5] \editorscompmeth;
\addplot[draw=black,fill=black!35, nodes near coords=MaxV] table[x index=0,y index=4] \editorscompmethtime;
\addplot[draw=black,fill=black!40, nodes near coords=PCF] table[x index=0,y index=5] \editorscompmethtime;
\end{axis}
 \end{tikzpicture}
 \end{footnotesize}
\caption{Comparison of methods w.r.t. running time (secs) for awk-sed. PCF (cycles=25). CVF (not shown) took at least an hour for 5 cycles.}
\label{fig:editors_cyclestime}
\end{figure*}

\begin{figure*}[th!]
\centering
\begin{footnotesize}
\begin{tikzpicture}
\begin{axis}[ybar,
        width=18cm,
        height=6cm,
        ymin=0,
        ymax=120,        
        ylabel={Median Score},
        xtick=data,
        xticklabels = {
cat,        
chown,
cp,
df,
echo,        
expand,
head,
ls
        },
        xticklabel style={yshift=-2ex},
        major x tick style = {opacity=0},
        minor x tick num = 1,
        minor tick length=0.2ex,
        every node near coord/.append style={
                anchor=east,
                rotate=90, 
        }
        ]
\addplot[draw=black,fill=black!10, nodes near coords=CIT2] table[x index=0,y index=1] \coreutilsmedian;        
\addplot[draw=black,fill=black!15, nodes near coords=CIT3] table[x index=0,y index=2] \coreutilsmedian;
\addplot[draw=black,fill=black!20, nodes near coords=MaxV] table[x index=0,y index=3] \coreutilsmedian;
\addplot[draw=black,fill=black!20, nodes near coords=PCF] table[x index=0,y index=4] \coreutilsmedian;
\end{axis}
 \end{tikzpicture}
 \end{footnotesize}
\caption{Comparison of CIT2, CIT3, MaxV, and PCF for median LCU score for cat-ls.}
\label{fig:coreutilsmedian}
\end{figure*}
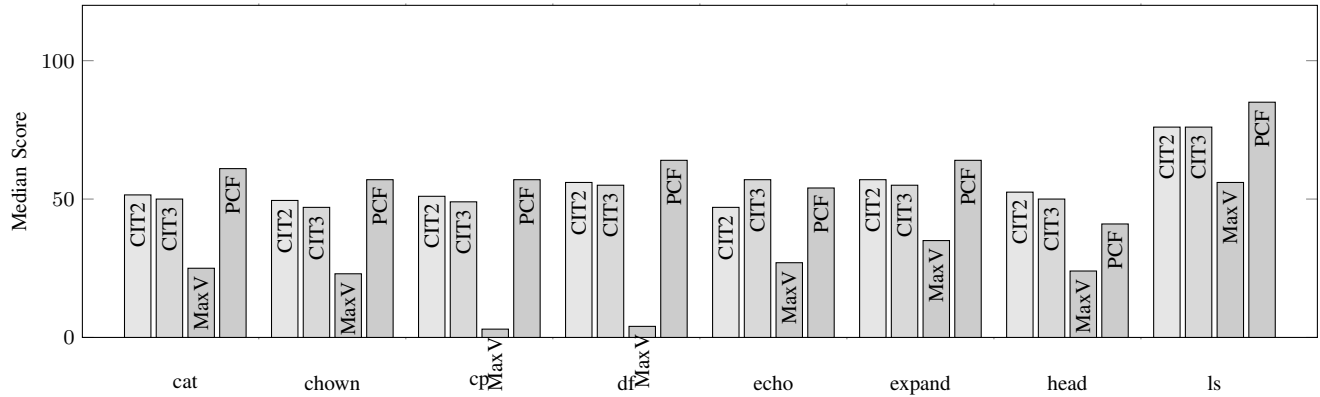

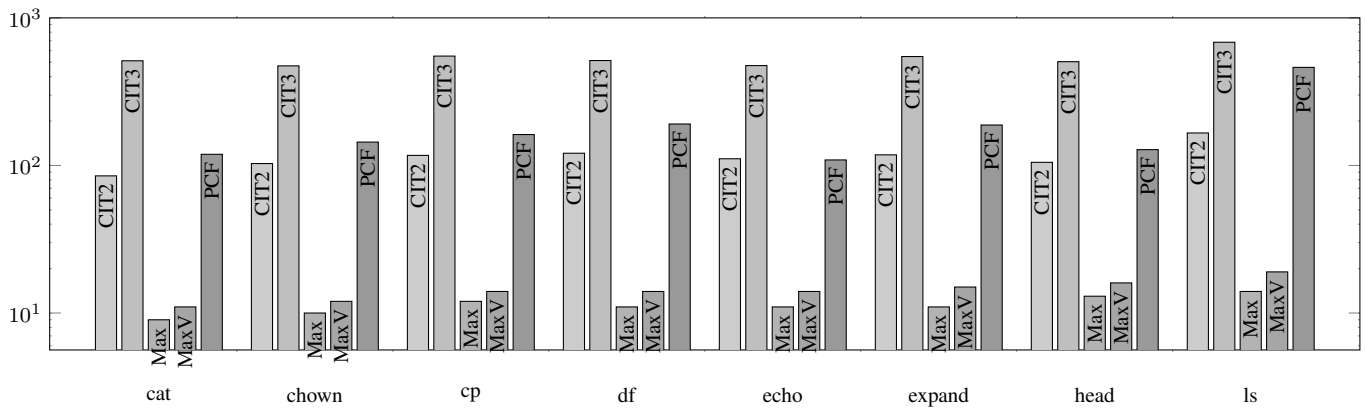
\begin{figure*}[th!]
\centering
\begin{footnotesize}
\begin{tikzpicture}
\begin{axis}[ybar,
        bar width=8pt,
        width=19cm,
        height=6cm,
        ymin=0,
        ymax=1000,        
        ymode=log,
        %minor xtick distance=0.5,
        xtick=data,
        xticklabels = {
cat, 
chown,
cp,
df,
echo,
expand,
head,
ls
        },
        xticklabel style={yshift=-2ex},
        major x tick style = {opacity=0},
        minor x tick num = 1,
        minor tick length=0.2ex,
        every node near coord/.append style={
                anchor=east,
                rotate=90
        }
        ]
%\addplot[draw=black,fill=black!10, nodes near coords=CVF] table[x index=0,y index=1] \editorscompmethtime1;        
%\addplot[draw=black,fill=black!15, nodes near coords=DC] table[x index=0,y index=2] \editorscompmethtime1;
\addplot[draw=black,fill=black!20, nodes near coords=CIT2] table[x index=0,y index=1] \coreutilscompmethtime;
\addplot[draw=black,fill=black!25, nodes near coords=CIT3] table[x index=0,y index=2] \coreutilscompmethtime;
\addplot[draw=black,fill=black!30, nodes near coords=Max] table[x index=0,y index=3] \coreutilscompmethtime;
%\addplot[draw=black,fill=black!20, nodes near coords=MSI] table[x index=0,y index=5] \editorscompmeth;
\addplot[draw=black,fill=black!35, nodes near coords={MaxV}] table[x index=0,y index=4] \coreutilscompmethtime;
\addplot[draw=black,fill=black!40, nodes near coords=PCF] table[x index=0,y index=5] \coreutilscompmethtime;
\end{axis}
 \end{tikzpicture}
 \end{footnotesize}
\caption{Comparison of methods w.r.t. running time (secs) for cat-ls. PCF (cycles=25). CVF (not shown) took at least an hour for 5 cycles.}
\label{fig:coreutils_cyclestime}
\end{figure*}

In terms of running time, as shown in Figures \ref{fig:editors_cyclestime} and \ref{fig:coreutils_cyclestime}, CVF is the slowest and takes several hours even when the number of cycles is set to 5. In some cases, it ran out of memory.
Maximal is the fastest followed by MaximalV while PCF is the 2nd slowest.
CIT3 is much slower than CIT2 as it generates more samples than CIT2.
When considering coverage and running time, Maximal excels in both. 
If the goal is to achieve 100\% code coverage then MaximalI must be the 
method of choice as it guarantees this.
To further evaluate the remaining methods, CIT2, CIT3, MaximalV, and PCF, we performed 
additional evaluation and computed the median LCU scores in the samples they generate.
As Figures \ref{fig:editorsmedian} and \ref{fig:coreutilsmedian} show,  
PCF provides the highest median scores indicating that if multiple configurations 
with high coverage are needed then the method of choice should be PCF, followed by CIT2.

\subsection{RQ4 on Bug Relevance of Code Metrics}

To answer {\bf RQ4: Do bug-relevant code metrics help generate configurations with relevant bugs?}, we used two code metrics: 1L-free, which refers to direct calls to the {\tt free} function, assigning a weight equal to the number of lines with calls to {\tt free} to the controlling presence condition, and 3L-free, which refers to direct and indirect calls to {\tt free} up to three levels of indirect calls, assigning the weights based on the number of direct and indirect calls.
Then we used AFL, a software fuzzing tool to fuzz components of the {\em editors}\footnote{Using AFL on {\em coreutils} components required extensions that go beyond the scope of this work.} subsystem that have calls to {\tt free}. We used the tests for these components in BusyBox and fuzzed each with different configurations. Table \ref{tab:busyboxfuzzing} shows the 1L-free and 3L-free scores of the configurations and how they are generated. 
As the table shows, crashes have been found when the configuration has a nonzero 1L-free or 
3L-free score while the case of both scores being zero did not yield any crashes. 
It is important to note that while crashes in awk were found by both the default configuration  and the maximal configuration, in the case of sed only the maximal configuration yielded 
crashes, all of which are related to the free calls\footnote{We are in the process of responsibly disclosing these crashes some of which have been tagged as exploitable by the afl-walk tool.}.
It thus appears that bug-relevant code metrics can be useful in generating 
configurations that help find bugs. 

Tables \ref{tab:free_methodcmp_editors} and \ref{tab:free_methodcmp_coreutils} show Maximal 
and its variants are the best among all the candidate approaches in terms of finding the maximum score configuration in the minimum number of steps, i.e., the number of configuration candidates generated, for the 1L-Free and 3L-Free code metrics, with Selective Presence Condition Fuzzing (SPCF) being the second best.

\pgfplotstableread{
0  61.2 66.8 67.8 67.8 67.6
1  66.4 70.8 72.4 72.0 72.4
2  63.0 70.2 71.8 72.0 71.4
3  61.4 66.8 66.2 65.4 66.8
4  64.8 69.8 71.0 70.2 70.0
5  63.6 70.2 69.8 70.4 70.8
6  61.4 65.0 66.2 65.4 66.6
7  64.8 69.8 69.4 70.0 70.2
    }\editorsdatamax

%\begin{comment}
\begin{figure*}[th!]
\centering
\begin{footnotesize}
\begin{tikzpicture}
\begin{axis}[ybar,
        bar width=8pt,
        width=18cm,
        height=6cm,
        ymin=0,
        ymax=80,        
        ylabel={Ave. Max. Score},
        xtick=data,
        xticklabels = {
awk,
cmp,
diff,
ed,
patch,
patch\_bbox,
patch\_toybox,
sed
        },
        xticklabel style={yshift=-2ex},
        major x tick style = {opacity=0},
        minor x tick num = 1,
        minor tick length=0.2ex,
        every node near coord/.append style={
                anchor=east,
                rotate=90
        }
        ]
\addplot[draw=black,fill=black!20, nodes near coords=c5] table[x index=0,y index=1] \editorsdatamax; 
\addplot[draw=black,fill=black!30, nodes near coords=c10] table[x index=0,y index=2] \editorsdatamax; 
\addplot[draw=black,fill=black!40, nodes near coords=c15] table[x index=0,y index=3] \editorsdatamax;
\addplot[draw=black,fill=black!50, nodes near coords=c20] table[x index=0,y index=4] \editorsdatamax;
\addplot[draw=black,fill=black!60, nodes near coords=c25] table[x index=0,y index=5] 
\editorsdatamax;
\end{axis}
 \end{tikzpicture}
 \end{footnotesize}
\caption{Maximum LCU score achieved by Presence Condition Fuzzing (PCF) for various number of fuzzing cycles for the editors components.}
\label{fig:editors_cyclesmax}
\end{figure*}
%\end{comment}

\pgfplotstableread{
0  64.6 71.8 72.8 72.8 72.6 
1  61.6 70.2 69.4 69.8 70.4 
2  64.4 72.4 72.6 73 74
3  65.6 73.4 76.8 74 76.4
4  61.8 67.2 67 67.4 66.8
5  72.4 78 83.8 81.6 82.8
6  66 72.8 74.2 73.4 74.6
7  89 97.4 94 104.0 105.2
    }\coreutilsdatamax

\begin{figure*}[th!]
\centering
\begin{footnotesize}
\begin{tikzpicture}
\begin{axis}[ybar,
        bar width=8pt,
        width=18cm,
        height=6cm,
        ymin=0,
        ymax=110,        
        ylabel={Ave. Max. Score},
        xtick=data,
        xticklabels = {
cat,
chown,
cp,
df,
echo,
expand,
head,
ls
        },
        xticklabel style={yshift=-2ex},
        major x tick style = {opacity=0},
        minor x tick num = 1,
        minor tick length=0.2ex,
        every node near coord/.append style={
                anchor=east,
                rotate=90
        }
        ]
\addplot[draw=black,fill=black!20, nodes near coords=c5] table[x index=0,y index=1] \coreutilsdatamax; 
\addplot[draw=black,fill=black!30, nodes near coords=c10] table[x index=0,y index=2] \coreutilsdatamax; 
\addplot[draw=black,fill=black!40, nodes near coords=c15] table[x index=0,y index=3] \coreutilsdatamax;
\addplot[draw=black,fill=black!50, nodes near coords=c20] table[x index=0,y index=4] \coreutilsdatamax;
\addplot[draw=black,fill=black!60, nodes near coords=c25] table[x index=0,y index=5] 
\coreutilsdatamax;
\end{axis}
 \end{tikzpicture}
 \end{footnotesize}
\caption{Maximum LCU score achieved by Presence Condition Fuzzing (PCF) for various number of fuzzing cycles for the coreutils components.}
\label{fig:coreutils_cyclesmax}
\end{figure*}

%awk-ed
\pgfplotstableread{
0 48 56 64 64 66 66 67.6 
1 47 62 70 70 72 74 72.4 
2 50 61 69 69 73 73 71.4  
3 47 55 63 63 65 65 66.8 
}\editorscompmetha

%patch_bbox-sed
 \pgfplotstableread{
0  48  59  67 67 71 71 70.8 
1  47  55  63 63 65 65 66.6 
2  48  59  67 67 71 71 70.0 
3  48  59  67 67 69 71 70.2 
}\editorscompmethb

\begin{figure*}[th!]
\centering
\begin{footnotesize}
\begin{tikzpicture}
\begin{axis}[ybar,
        bar width=8pt,
        width=18cm,
        height=6cm,
        ymin=0,
        ymax=75,        
        ylabel={Ave. Max. Score},
        xtick=data,
        xticklabels = {
awk,        
cmp,
diff,
ed
        },
        xticklabel style={yshift=-2ex},
        major x tick style = {opacity=0},
        minor x tick num = 1,
        minor tick length=0.2ex,
        every node near coord/.append style={
                anchor=east,
                rotate=90
        }
        ]
\addplot[draw=black,fill=black!10, nodes near coords=CVF] table[x index=0,y index=1] \editorscompmetha;        
\addplot[draw=black,fill=black!15, nodes near coords=DefConf] table[x index=0,y index=2] \editorscompmetha;
\addplot[draw=black,fill=black!20, nodes near coords=CIT2] table[x index=0,y index=3] \editorscompmetha;
\addplot[draw=black,fill=black!25, nodes near coords=CIT3] table[x index=0,y index=4] \editorscompmetha;
\addplot[draw=black,fill=black!30, nodes near coords=Max] table[x index=0,y index=5] \editorscompmetha;
%\addplot[draw=black,fill=black!20, nodes near coords=MSI] table[x index=0,y index=5] \editorscompmeth;
\addplot[draw=black,fill=black!35, nodes near coords=MaxV] table[x index=0,y index=6] \editorscompmetha;
\addplot[draw=black,fill=black!40, nodes near coords=PCF] table[x index=0,y index=7] \editorscompmetha;
\end{axis}
 \end{tikzpicture}
 \end{footnotesize}
\caption{Comparison of methods w.r.t. LCU max score for awk-ed. PCF (cycles=25). CVF (cycles=5).}
\label{fig:editors1_cyclesmax}
\end{figure*}

% medians of cit2, cit3, maxV, PCF
% awk, cmp, diff,ed, 
% patch\_box, patch\_toybox, patch, sed
\pgfplotstableread{
0 47 47 26 55.8
1 52.5 51.5 29 53.6
2 50 48 25 58.8
3 46 46 25 53.2
4 49.5 48 22 57.2 
5 46 46 25 53.6
6 49.5 48 22 55.6
7 49.5 48.5 27 56
}\editorsmedian

%\pgfplotstableread{
%0 3 172 19.57 28.07 512.31 
%1 3 172 20.56 23.85 496.67  
%2 3 172 22.63 29.06 543.95   
%3 3 172 16.50 41.73 435.97
%4 3 172 20.40 36.35 468.83 
%5 3 172 25.41 41.86 354.84  
%6 3 172 18.38 36.34 354.78   
%7 3 172 18.97 60.89 370.74  
%}\editorscompmethtime

\pgfplotstableread{
0 122 491 6 12  114 
1 151 519 9 11 115  
2 97 520 10 12 198   
3 91 483 10 12 146
4 85 538 9 10 141 
5 78 499 12 14 155  
6 87 507 11 13 177   
7 83 528 12 15 169  
}\editorscompmethtime

\pgfplotstableread{
0  51 62 70  70 74 74 72.06 
1  50 59 67 67 71 71 70.4
2  53 61 69 69 73 73 74
3  56 60 78 78 70 74 76.4
}\coreutilscompmetha

\pgfplotstableread{
0  48 56 64 64 66 66 66.8
1  57  62 81 81 78 78 82.8
2  50  63 71 71 70 70 74.6
3  68  89 106 108 94 98 105.2
}\coreutilscompmethb

% median score for CIT2, CIT3, MaxV, PCF
%cat, chown, cp, df,
%echo, expand, head, ls
\pgfplotstableread{
0 51.5  50 25 61
1 49.5 47 23 57
2 51 49 3 57 
3 56 55 4 64 
4 47 57 27 54
5 57 55 35 64
6 52.5 50 24 41
7 76 76 56 85
}\coreutilsmedian

%\pgfplotstableread{
%0 3 172 16.04 58.53 299.07
%1 3 172 18.20 48.47 203.55
%2 3 172 18.19 75.49 258.51
%3 3 172 17.35 56.18 365.39
%4 3 172 17.53 32.80 193.30
%5 3 172 18.82 70.01 416.08
%6 3 172 19.75 65.99 125.03
%7 3 172 37.29 144.38 664.22
%}\coreutilscompmethtime

\pgfplotstableread{
0 85 512 9 11 119
1 103 473 10 12 144
2 117 551 12 14 162
3 121 514 11 14 191
4 111 475 11 14 109
5 118 547 11 15 188
6 105 505 13 16 128
7 166 684 14 19 462
}\coreutilscompmethtime

\section{Conclusions}
\label{sec:conclusions}

Our experiments with CONFIZZ provide examples of how it can reveal valid configurations of interest that may have been unknown to testers for their focused attention and testing.  Especially for large configurable systems, CONFIZZ's results can offer testers an additional perspective, and perhaps additional understanding, regarding configurations that merit their attention. 

Projects typically want to test certain sets of configurations, such as those involving critical behaviors,  more thoroughly than other configurations. CONFIZZ enables this customized focus via  maximal configurations and code metrics, which can be defined to target specific bugs.   Such flexibility can be especially important in budget and/or time constrained projects where test time is a costly or scarce resource. 

In this work we presented a new
configuration-generation framework, CONFIZZ, that uses  MaxSAT-based statement
coverage sampling We described how CONFIZZ can both generate a minimum number of configurations that maximize the user-selected code metric function, and can perform presence condition fuzzing.  Results from evaluation on BusyBox show that MaxSAT-based configuration generation 
can help achieve full statement coverage and high coverage for other code metrics such as those  related to memory deallocation. Our results show that bug relevant code metrics can 
guide configuration testing in bug finding. We also show that Presence Condition Fuzzing is 
a good alternative to MaxSAT-based configuration generation.

\section{Acknowledgements}
This work has been partially funded by NSF awards CCF-2211588 and CCF-2211589.
%\clearpage
\bibliographystyle{IEEEtran}
\bibliography{feature.bib}

\end{document}